\documentstyle[12pt,version2,aps,preprint]{revtex}

\def\Delq{\Delta_{\sbox{q}}}
\def\Delt{\Delta_{\sbox{t}}}

\def\ln{\mbox{ln}}
\def\sztot{S_z^{\sbox{tot}}}
\def\sbox#1{\mbox{\scriptsize #1}}

\def\v#1{\mbox{\boldmath$#1$}}

\def\simgeq{\mbox{\raisebox{-1.0ex}{$\stackrel{>}{\sim}$}}}

\begin{document}
\begin{title}
{\bf Density Matrix Renormalization Group Study\\ of the Spin 1/2 Heisenberg
Ladder with\\ Antiferromagnetic Legs and Ferromagnetic Rungs
}
\end{title}
\author{
 Kazuo HIDA}

\begin{instit}

\it
Department of Physics, Faculty of Science, \\ Saitama University, Urawa,
Saitama 338
\end{instit}

\begin{center}
(Received\hspace{4cm})
\end{center}

The ground state and low lying excitation of the spin 1/2 Heisenberg ladder
with antiferromagnetic leg ($J$) and ferromagnetic rung ($-\lambda J, \lambda
>0$) interaction is studied by means of the density matrix renormalization
group method. It is found that the state remains in the Haldane phase even for
small $\lambda \sim 0.02$ suggesting the continuous transition to the gapless
phase at $\lambda = 0$. The critical behavior for small $\lambda$ is studied by
the finite size scaling analysis. The result is consistent with the recent
field theoretical prediction.

\vspace{2mm}
\noindent
Keywords:  Heisenberg  ladder,  Haldane gap, Density matrix renormalization
group

\noindent
e-mail: hida@th.phy.saitama-u.ac.jp

\sloppy
\maketitle

\section{Introduction}

Recently, the spin-1/2 Heisenberg ladder model has been studied
extensively\cite{kh1,kh2,wnt1,drs1,bdrs1,grs1,acm1,br1,ttw1,
wns1,sm1,sm2,ts1,w1,nishi1,at1,naru1} motivated by its relevance to high $T_c$
superconductivity and Haldane gap problem.\cite{hd1,hd2,ia1,tk1} The former is
concerned with the fully antiferromagnetic ladder, while the latter is related
to the ladder with ferromagnetic rung and antiferromagnetic leg interactions.
In the present paper, we concentrate on the latter model.

Several years ago, the present author studied this model using the projctor
Monte Carlo method.\cite{kh1} It has been further studied by the non-linear
$\sigma$ model,\cite{kh2} the bosonization method\cite{wnt1,sm1,sm2,ts1} and
exact diagonalization.\cite{w1} For small interchain coupling, however, their
conclusions are still controversial. The bosonization calculation by Watanabe,
Nomura and Takada\cite{wnt1} support the Kosterlitz-Thouless (KT) type behavior
with the energy gap proportional to $\exp(-\mbox{const.}/\sqrt{\lambda})$, and
the nonlinear $\sigma$ model analysis\cite{kh2} predicts the possiblity of the
KT type transition to the gapless state at nonzero positive $\lambda$. Although
the projector Monte Carlo simulation\cite{kh1} seems to support one of these
two kinds of possibility, the data are rather scattered. On the other hand, the
non-Abelian bosonization studies by Strong and Millis\cite{sm1,sm2} suggest the
continuous transition at $\lambda = 0$ where energy gap is proportional to
$\lambda$.
Using the same method, Totsuka and Suzuki\cite{ts1} predict the presence of the
logarithmic correction to the gap as $\lambda (\ln(1/\lambda))^{1/2}$. The
exact diagonalization study of this model has been also performed by
Watanabe.\cite{w1} Nishiyama {\it et al}.\cite{nishi1} and Terai\cite{at1} made
the perturbational approach from $\lambda =0$ for the finite size system. The
density matrix renormalization group (DMRG)\cite{rw1,rw2} study is also carried
out by Narushima {\it et al}.\cite{naru1} These works support the continuous
transition at $\lambda = 0$ with the energy gap exponent $\nu \simeq 1$ with
possible logarithmic correction.  It is the purpose of the present work to
investigete this problem further analyzing the DMRG data by the finite size
scaling method.

In the next section, we explain the model and the method of the DMRG
calculation. The results are analyzed in sections 3 and 4. The last section is
devoted to summary and discussion.

\section {Heisenberg Ladder}

We consider the spin 1/2 Heisenberg ladder consisting of two coupled Heisnberg
chains represented by the following Hamiltonian $H$.
\begin{equation}
\label{eq:ham1}
H =2J\sum_{i=1}^{N}\left[\v{S}^A_{i} \v{S}^A_{i+1}+\v{S}^B_{i} \v{S}^B_{i+1} -
\lambda\v{S}^A_{i} \v{S}^B_{i} \right],\ \ \ (J, \lambda > 0)
\end{equation}
where $\v{S}^{\alpha}_i (\alpha = A \mbox{ or } B )$ is the spin operator with
spin 1/2. The length of the ladder is denoted by $N$.  For $\lambda \rightarrow
\infty$, the spins $\v{S}^A_{i}$ and $\v{S}^B_{i}$ form a local triplet and
this model reduces to the spin-1 Heisenberg chain. On the other hand, this
model is decoupled into two spin 1/2 antiferromagnetic Heisenberg chains for
$\lambda=0$.

In the DMRG transform,\cite{rw1,rw2} the number $m$ of the states retained in
the left and right half systems ranges from 80 to 140. We use the mixed
algorithm of finite size and infinite size method. Namely, during the increase
of the system size by the infinite size algorithm, we sometimes improve the
accuracy by moving the boundary between the left and right half systems until
the eigenvalues converge keeping the system size constant. In order to save the
computational time, however, we only move the boundary among the sites which
are added after the last finite size iteration. The data are taken only for the
sizes for which the finite size iteration is made. For $m=140$, only the
infinite size algorithm is used unless specifically mentioned. On each step of
the infinite size iteration, 4 spins are added. In the following, we only
consider the case of even $N$.

In the calculation of the low lying excitation energy eigenvalues, we have
constructed the density matrices of the left and right half systems from the
lowest 3 states of the whole system with fixed $\sztot$(= $z$-component of the
total spin). Let us denote the energy eigenvalue of the $n$-th excited state
with $\sztot$ by $E(n,\sztot)$ where $n=0$ corresponds to the ground state with
given value of $\sztot$. For large $N$, the calculated energy gap depend on
$m$. We assume that the finite $m$ correction behaves as $1/m^2$ for large $m$
as pointed out by Narushima {\it et al}.\cite{naru1} Taking this into account,
we tried two different ways of $m$-extrapolation. First, the data are
extrapolated from $m = 100 , 120$ and 140 linearly to $1/m^2$ by the least
squares fit. Secondly, $1/m^3$-correction is included using the data for $m =
80, 100, 120$ and 140. The data points are the average of these two
extrapolated values and the error bars are estimated from the difference
between them.

\section {Kennedy Triplet}

One of the most remarkable features of the Haldane phase is the presence of the
four-fold quasi-degeneracy of the ground state and low lying states\cite{tk1}
in the open boundary system. In analogy with the $S=1$ antiferromagnetic
Heisenberg chain, the ground state is a singlet and the three lowest excited
states form a triplet called Kennedy triplet\cite{tk1} if the ladder length $N$
is even.

 Fig. 1 shows the semi-log plot of the size dependence of the excitation energy
$\Delta_K = E(1,0)-E(0,0)$ of the lowest excited states with $\sztot =0$
extrapolated to $m \rightarrow \infty$. The convergence of the numerical
diagonalization procedure becomes poor for some points due to the reason
explained in the next section. These data are excluded from the analysis. For
large $N$, the excitation energy vanishes exponentially with the increase of
the system size within the error bars confirming that the system is in the
Haldane phase down to $\lambda \sim 0.02$. This implies that the continuous
transition to the gapless phase takes place at $\lambda=0$. The same conclusion
is also obtained by Narushima {\it et al}.\cite{naru1}

\section {Haldane Gap}

Taking into account the presence of the Kennedy triplet, the Haldane gap  may
be estimated by extrapolating the difference
 $\Delta E_0 = E(2,0)-E(1,0)$\cite{w1} to $N \rightarrow \infty$. The system
size dependence of $\Delta E_0$ is shown in Fig. 2 for $\lambda= 0.02, 0.05,
0.1, 0.15 \mbox{ and } 0.2$ and $4 \leq N \leq 102$. The second excitation gap
$\Delta E_1 =E(3,0)-E(1,0)$ is also shown for small systems ($N \leq 12$) which
can be diagonalized exactly. From this figure, we find that the gap $\Delta
E_0$ has a maximum at $N =N_{\sbox{max}}(\lambda)$ which increases with the
decrease of $\lambda$.

This maximum results from the level crossing which takes place in the finite
size system at $\lambda =\lambda_{\sbox{c}}(N)$. This level crossing is already
pointed out by Watanabe\cite{w1} and the physical picture of these excitations
are also given. Roughly speaking, we may regard $N_{\sbox{max}}(\lambda)$ as
the inverse function of $\lambda_c(N)$, although this is not well-defined in
the mathematical sense because $N$ is a discrete variable.

 The gap $\Delta E_0$  for $N > N_{\sbox{max}}(\lambda)$ continue to the second
excitation gap $\Delta E_1$ for  $N < N_{\sbox{max}}(\lambda)$ as shown in Fig.
2. Let us define the gap $\Delq$ by $\Delta E_0$ for  $N > N_{\sbox{max}}$ and
$\Delta E_1$ for $N < N_{\sbox{max}}$, because this state corresponds to the
quintuplet excitation.\cite{w1} In the thermodynamic limit $N \rightarrow
\infty$, $\Delq$ corresponds to the Haldane gap. Similarly, second triplet
excitation gap
 $\Delt$ is defined by $\Delta E_1$ for  $N > N_{\sbox{max}}$ and   $\Delta
E_0$ for $N < N_{\sbox{max}}$, although we did not calculate $\Delta E_1$ for
$N > N_{\sbox{max}}$ because the accuracy of the higher excitation energy
becomes worse in the DMRG method. Nevertheless, we have shown
$\Delta_{\sbox{t}}$ for $\lambda=0.05$ for large $N$. These eigenvalues are
obtained as the second excited state in the DMRG caluculation. However, we
interprete these values as  $\Delta_{\sbox{t}}$ instead of $\Delta_{\sbox{q}}$,
because it is on the line smoothly extrapolated from $\Delta_{\sbox{t}}$. We
expect that this occurs because the states important for the quintuplet states
are already discarded during the iteration for $N < N_{\sbox{max}}$. Due to the
same reason, the gap $\Delq$ is less accurate than $\Delt$ for large $N$.  This
makes difficult to determine the precise value of the exponent $\nu$ for
$\Delta_{\sbox{q}}$.

We therefore start with the analysis of the system size dependence of $\Delt$
by the finite size scaling method. In this analysis, we use only the data
obtained after the finite size iteration even for $m=140$. According to Strong
and Millis\cite{sm1,sm2} and Totsuka and Suzuki,\cite{ts1} the field
theoretical calculation using the non-Abelian bosonization technique yields
$\nu=1$. Totsuka and Suzuki\cite{ts1} also noted the presence of the
logarithmic correction to the gap as $\lambda (\ln(1/\lambda))^{1/2}$.
Expecting the similar behavior for $\Delt$, let us assume the finite size
scaling formula for $\Delta_{\sbox{t}}$ as,

\begin{equation}
\label{myfit}
N\Delta_{\sbox{t}}(N) = f(N(\ln N)^{\alpha} \lambda^{\nu})
\end{equation}
This scaling formula is also employed by Terai.\cite{at1} As discussed by
Spronken {\it et al}. \cite{sp1}, this scaling relation implies that the gap
$\Delta_{\sbox{t}}$ of the infinite system behaves as $\Delta_{\sbox{t}}\sim
\lambda^{\nu} (\mbox{ln}(1/\lambda))^{\alpha}$. We have tried to fix both
$\alpha$ and $\nu$ by the least squares fit to the universal curve $f$ using
the data for $\lambda = 0.02, 0.05, 0.1 $ and 0.2 and $4 \leq N \leq 102$. The
data with large error bars ($ \delta (N \Delt) \simgeq 1$) are excluded from
the analysis. However, the least squares deviation is rather insensitive to the
value of $\alpha$ while the optimum value of $\nu$ remains around unity
irrespective of the choice of $\alpha$.  For example, Fig. 3 shows the plot
with $\nu=0.88$ and $\alpha=0$  which is the optimum choice with $\alpha=0$.
 Fig. 4 shows the fit with $\alpha = 0.4$ and $\nu=1$ which is the optimum
choice with $\nu=1$. In both cases, the data $N\Delt$ excellently fits the
single universal curve $f(x) (x=N(\ln N)^{\alpha} \lambda^{\nu})$. Although the
available value of $x$ for $\Delt$ is limited, $f(x)$ behaves linearly for
large $x$ within the present data. The scaling plot of $\Delq$ is also shown.
For large $N$, it also follows another universal curve which also tends to a
linear function of $x$ with the same scaling exponent fairly well. This
suggests that the critical behavior of $\Delq$ is also governed by the same
exponents as those for $\Delt$.

We can obtain almost equally good fit using other sets of values of $\alpha$
and $\nu$ as far as $\nu \sim 1$ both for $\Delt$ and $\Delq$. This clearly
excludes the possibility of KT type behavior. However, this in turn means that
it is difficult to determine the value of $\alpha$ solely from the numerical
data. Therefore, taking into account the estimations from other
works\cite{sm1,sm2,ts1,nishi1,at1,naru1} for the exponents of the Haldane gap
and assuming that $\Delq$ and $\Delt$ has the same exponents as discussed
above, we fix the value of $\nu$ to unity and search for the optimum value of
$\alpha$. In this way, we find $\alpha = 0.4 \pm 0.07$.  The error of $\alpha$
is estimated from that of $\Delt$ due to $m$-extrapolation. This value is close
to the analytical estimation $\alpha = 0.5$ by Totsuka and Suzuki\cite{ts1} and
perturbational estimation of \cite{at1}.

 The finite size scaling plot in Fig. 4 also suggests that the level crossing
point is also scale invariant as $\lambda_c(N) N( \ln N)^{\alpha} =
\mbox{const.} \sim 5.2$. This implies that $\lambda_c$ tends to 0 as $N
\rightarrow \infty$, confirming that the gap $\Delq$ remains the lowest gap
down to $\lambda =0$ in the thermodynamic limit.

\section {Summary and Discussion.}

We have studied the low energy excitation of the spin-1/2 Heisenberg ladder
with ferromagnetic rungs and antiferromagnetic legs using the DMRG method. It
is found that the ground state remains in the Haldane phase down to $\lambda =
0.02$ suggesting the continous transition to the gapless phase at $\lambda =
0$.
 The exponent $\nu$ of the second triplet gap $\Delt$ is estimated to be around
unity using the finite size scaling analysis. Fixing  $\nu=1$, we find the
logarithmic correction as  $\Delt \sim \lambda (\ln (1/\lambda))^{\alpha}$ with
$\alpha =0.4 \pm 0.07$. Although the numerically obtained value of $\Delq$ is
not enough accurate to determine the exponents independently from $\Delt$, it
is expected $\Delq$ also shows the same critical behavior taking into account
the fact that $\Delq$ also fits to the universal curve fairly well with the
same scaling within the error bars. It is also shown that $\Delq$ remains the
lowest gap (Haldane gap) down to $\lambda \rightarrow 0$ in the thermodynamic
limit.

Several years ago, the present author proposed the possibility of the
transition at finite $\lambda$ based on the projector Monte Carlo
simulation\cite{kh1} and the mapping onto the nonlinear $\sigma$
model.\cite{kh2} However, the Monte Carlo data are rather scattered and the
nonlinear $\sigma$ model analysis is based on the semiclassical approximation
and the artificially introduced anisotropy. Therefore, we believe the present
result is much more reliable. Actually, the non-Abelian bosonization analysis
employed by Strong and Millis\cite{sm1,sm2} and Totsuka and Suzuki,\cite{ts1}
which leads to the results consistent with our calculation, may be regarded as
the refined version of the nonlinear $\sigma$ model analysis suited for the
isotropic case.

\acknowledgements

The author is indebted to H. Nishimori for TITPACK ver.2 for the
diagonalization of spin 1/2 system. He also thanks N. Hatano, A. Terai and H.
Watanabe for comments and discussion. This work is partly supported by the
Grand-in-Aid for Scientific Research from the Ministry of Education, Science
and Culture. The numerical calculation has been performed using the FACOM
VPP500 at the Supercomputer Center, Institute for Solid State Physics,
University of Tokyo and the HITAC S820/15 at the Information Processing Center,
Saitama University.


\newpage
\figure{The system size dependence of the energy gap  $\Delta_K$ between the
ground state and the Kennedy triplet for $\lambda=0.02, 0.05, 0.1, 0.15$ and
0.2.}

\figure{The system size dependence of the energy gap $\Delta E_0$ (open
symbols) and $\Delta E_1$ (filled symbols) for $\lambda=0.02, 0.05, 0.1, 0.15$
and 0.2.}

\figure{The finite size scaling plot with $\nu=0.88$ of the energy gap
$\Delta_{\sbox{t}}$ (open symbols) and $\Delta_{\sbox{q}}$ (filled symbols) for
$\lambda=0.02, 0.05, 0.1, 0.15$ and 0.2.}

\figure{The finite size scaling plot with $\nu=1$ and $\alpha=0.4$ of the
energy gap $\Delta_{\sbox{t}}$ (open symbols) and $\Delta_{\sbox{q}}$ (filled
symbols) for $\lambda=0.02, 0.05, 0.1, 0.15$ and 0.2. }

\vspace{5mm}

\begin{thebibliography}{99}
\bibitem{kh1}   K. Hida: J. Phys. Soc. Jpn. {\bf 60} (1991) 1347.
\bibitem{kh2}   K. Hida: J. Phys. Soc. Jpn. {\bf 60} (1991) 1939.

\bibitem{wnt1}  H. Watanabe, K. Nomura and S. Takada: J. Phys. Soc. Jpn. {\bf
62} (1993) 2845.

\bibitem{drs1}  E. Dagotto, J. Riera and D. Scalapino: Phys. Rev. {\bf B45}
(1992) 5744.
\bibitem{bdrs1}  T. Barnes, E. Dagotto, J. Riera and E.S. Swanson: Phys. Rev.
{\bf B47} (1993) 3196.
\bibitem{grs1}  Sudha Gopalan, T.M. Rice and M. Sigrist: Phys. Rev. {\bf B49}
(1994) 8901.
\bibitem{acm1}  M. Azzouz, L. Chen and S. Moukouri: Phys. Rev. {\bf B50} (1994)
6233.
\bibitem{br1}  T. Barnes and J. Riera: Phys. Rev. {\bf B50} (1994) 6817.
\bibitem{ttw1}   M. Troyer, H. Tsunetsugu and D. W\"urtz: Phys. Rev. {\bf B50}
13515 (1994).
\bibitem{wns1}   S. R. White, R. M. Noack and D. J. Scalapino: Phys. Rev. Lett.
{\bf 73} 886 (1994).
\bibitem{sm1}   S. P. Strong and A. J. Millis: Phys. Rev. Lett. {\bf 69} 2419
(1992).
\bibitem{sm2}   S. P. Strong and A. J. Millis: Phys. Rev. {\bf B50} 9911
(1994).
\bibitem{ts1}  K. Totsuka and M. Suzuki: J. Phys. Condens. Matter {\bf 7} 6079
(1995).
\bibitem{w1}    H. Watanabe: Phys. Rev. {\bf B50} 13442 (1994).
\bibitem{nishi1}  Y. Nishiyama, N. Hatano and M. Suzuki: J. Phys.Soc. Jpn {\bf
64} 1967 (1995).
\bibitem{at1}  A. Terai: private communication.
\bibitem{naru1}  T. Narushima, T. Nakamura and S. Takada: preprint
(1995).\bibitem{hd1}   F.D.M. Haldane: Phys. Lett. {\bf 93A} 464 (1983).
\bibitem{hd2}   F.D.M. Haldane: Phys. Rev. Lett. {\bf 50} 1153 (1983).
\bibitem{ia1}	I. Affleck: J. Phys. Condens. Matter. {\bf 1} 3047 (1989) and
references therein.



\bibitem{tk1}   T. Kennedy: J. Phys. Condens. Matter. {\bf 2} (1990) 5737.

\bibitem{rw1}   S. R. White: Phys. Rev. Lett. {\bf 69} 2863 (1992).

\bibitem{rw2}   S. R. White: Phys. Rev. {\bf B48} 10345 (1993).


\bibitem{sp1}   G. Spronken, B. Fourcade and Y. L\'epine:  Phys. Rev. {\bf B33}
1886 (1986).

\end{thebibliography}
\end{document}